\documentclass[showpacs,prl,aps, twocolumn]{revtex4}
\usepackage{graphicx}% Include figure files
\usepackage{color}
\usepackage{dcolumn}% Align table columns on decimal point
\usepackage{bm}% bold math
\def\ph2{{\it p}-H$_2$}
\def\gapx{\lower 2pt \hbox{$\buildrel>\over{\scriptstyle{\sim}}$\ }}
\def\lapx{\lower 2pt \hbox{$\buildrel<\over{\scriptstyle{\sim}}$\ }}
\def\Am3{\AA$^{-3}$}
\begin{document}
\title{Local Superfluidity of  Parahydrogen Clusters}

\author{Fabio Mezzacapo}
\author{Massimo  Boninsegni}

\affiliation{Department of Physics, University of Alberta, Edmonton, Alberta, Canada T6G 2G7 }
\date{\today}

\begin{abstract}
We study by Quantum Monte Carlo simulations the local superfluid response of small (up to 27 molecules) {para}hydrogen clusters, down to temperatures as  low as 0.05 K. We show that at low temperature superfluidity is {\it not} confined at the surface of the clusters, as recently claimed by Khairallah {\it et al.} [Phys. Rev. Lett. {\bf 98}, 183401 (2007)]. Rather, even clusters with a pronounced shell structure are essentially {\it uniformly} superfluid. Superfluidity occurs as a result of long exchange cycles involving all molecules.
\end{abstract}

\pacs{67.90.+z, 61.25.Em}

\maketitle
Small clusters of   {para}hydrogen (\ph2) molecules are believed to remain liquidlike at significantly lower temperature than that at which bulk \ph2 crystallizes, rendering the observation of superfluidity (SF) possible \cite{sindzingre,noi06,noi07a,buffoni}. Indeed, some experimental evidence of SF has been obtained  for  \ph2 clusters of approximately 15 molecules, at a temperature $T$=0.15  K  \cite{grebenev00}.  Those findings prompted an ongoing theoretical research effort, aimed at characterizing structural and superfluid properties of \ph2 clusters, both pristine and doped with impurities \cite{noi06,noi07a, buffoni,kwon02_05,paesani05,toennies,saverio05,guardiola06,roy06,noi07b}. The theoretical method of choice has been computer simulations based on Quantum Monte Carlo (QMC) techniques, at zero or  finite temperature, which yield essentially exact numerical estimates for Bose systems.  

Although numerical discrepancies exist among the various works, attributable in some cases to the use of different interaction potentials, a finite global superfluid response has been generally predicted for pure \ph2 clusters of size at least up to $N$=27 molecules, at $T \le$ 1 K; some clusters (of size 22 $\leq N \leq$ 30) turn superfluid as they undergo {\it quantum melting}, going from solid- to liquidlike when $T$ is lowered, purely as a result of quantum exchanges; the solid bulk phase is found to emerge non-monotonically  for increasing cluster size \cite {noi06, noi07a}.

 One interesting open issue is how the superfluid response is distributed across the droplet.  Because a cluster is a finite non-homogeneous system, one may meaningfully pose the question of whether the decoupling from an externally applied rotation characterizing a superfluid is greater in some parts thereof (e.g., the surface). The relevance of this aspect goes obviously beyond the physics of \ph2 clusters,  as it impacts the general theoretical understanding of SF, and its connection with Bose condensation.

In a recent paper, Khairallah {\it et al.} utilized the radial density profile of particles involved in permutation cycles to study the local superfluid density of \ph2 droplets. Based on their results, they concluded that the superfluid response  is largely confined at the surface of these small clusters, mainly arising from exchange cycles involving loosely bound surface molecules \cite{buffoni}.

Aside from the difficulty of providing a meaningful definition of  ``surface'' 
 of clusters of such small size, the physical picture proposed in Ref. \cite{buffoni} is at odds with recent simulation work by us, on pure and isotopically doped \ph2 clusters, which has yielded indirect evidence that their superfluid response correlates with  permutational exchange cycles comprising molecules in both the inner and the outer shell \cite{noi06,noi07a,noi07b}. Indeed, even clusters that  display a coexistence of liquid- and solidlike behavior at $T\sim$ 1 K [e.g., (\ph2)$_{23}$] are essentially {\it entirely} superfluid in the liquidlike ``phase" \cite{noi06,noi07a}, an observation which 
raises doubts about the contention that SF may be confined at the surface of the cluster.

In order to provide an independent theoretical check of the predictions of Ref. \cite{buffoni}, we have undertaken in this work a systematic study of the local superfluid properties of \ph2 clusters at low temperature, using QMC simulations with a recently introduced, rigorous local superfluid density estimator, consistent with the basic two-fluid model \cite{whaley}.
Specifically we have computed the  local superfluid response of pristine \ph2 clusters of up to 27 molecules in the temperature range 0.05 K $\le T\le$ 3 K, and examined its connection to global properties previously discussed in Refs. \cite {noi06, noi07a}.

Our results show that, contrary to what stated in  Ref. \cite{buffoni}, SF in these clusters is {\it not} localized at the surface; rather, in the low temperature limit clusters are {\it uniformly} superfluid; this statement applies not only to the more liquidlike ones (i.e., those with fewer than $\sim 20$ molecules), but also to those displaying a more pronounced shell structure, which softens at low temperature due to quantum melting, in turn underlain by exchange cycles involving molecules in both the first and second shells. Based on these findings we contend that the physical picture proposed in Ref. \cite {buffoni} of SF as arising from ``loosely bound  surface molecules" does not apply. Indeed, we have observed no clusters with a rigid core and a superfluid outer shell. Instead, SF crucially depends on the onset of long exchange cycles involving {\it all} molecules, not just those on the surface.

We adopted the usual microscopic model for  our system of interest, namely a collection of $N$ \ph2 molecules regarded as Bose particles of spin $S$=0, interacting via a central pair potential. We used the most commonly adopted model  interaction for \ph2 clusters, namely the Silvera-Goldman potential \cite{SilveraandGoldman}, in all the calculations for which results are shown here.  We studied this system by QMC simulations at finite temperature based on the continuous-space worm algorithm \cite{MBworm, worm2}, which gives us access to temperatures as low as 0.05 K; technical details of our calculations are the same of Ref. \cite{noi07a}. Profiles of radial superfluid density were computed using the  local estimator proposed in Ref. \cite{whaley}; specifically, the local superfluid response $\rho_S(r)$ is expressed as follows:
\begin{equation}
\frac{\rho_S(r)}{\rho(r)}=\frac{4m^2\langle AA(r)\rangle}{\beta\hbar^2 I_C(r)}
\label{est}
\end{equation}
where $m$ is the mass of a \ph2 molecule, $\beta = 1/T$,  $\rho(r)$ is the local density and  $\langle\cdots\rangle$ stands for the thermal average. $A$ is  the total area swept by the many-particle paths projected onto a plane perpendicular to one of the three equivalent rotation axes, whereas $A(r)$ and $I_C(r)$ are, respectively, the contributions to the total area $A$ and to the classical moment of inertia $I_C$ from a spherical shell of radius $r$ centered at the center of mass of the cluster. The above estimator  is obtained from the definition of normal fraction as proportional to the linear response of a system to an externally applied rotation. The response of the system as a whole is decomposed in separate contributions, which provide local information.
\begin{figure}
\centerline{\includegraphics[scale=0.55,angle=-90]{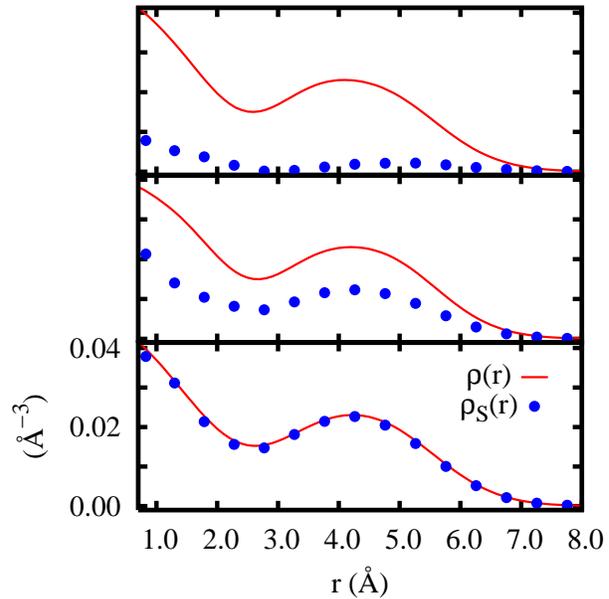}}
\caption{(color online) Profiles of total [$\rho(r)$] and superfluid [$\rho_S(r)$] density computed with respect to the center of mass of the cluster (\ph2)$_{18}$ at $T$ = 1 K (lower panel), $T$ = 2 K (middle panel) and $T$ = 3 K (upper panel). }
\label{fig:18}
\end{figure}

Figure \ref{fig:18} shows profiles of  total [$\rho(r)$] and superfluid [$\rho_S(r)$]  radial densities, computed with respect to the center of mass of the system,   for a cluster comprising $N$=18 \ph2 molecules. At $T$ = 1 K the cluster  is entirely superfluid, within the statistical uncertainty of our calculation \cite{noi06, noi07a}. As shown in Fig. \ref{fig:18}, the superfluid fraction is 100\% everywhere in the cluster, with no sign of weakening near the center. At $T$ = 2 K,   the total superfluid fraction of the cluster decreases to $\sim$ 55\%; notably, however, the superfluid density remains finite throughout the cluster, even in its inner region. Upon increasing $T$ to 3 K, the superfluid fraction drops to about 14\% and  $\rho_S(r)$ is correspondingly uniformly depressed throughout the whole system, almost completely  in correspondence of the total density minimum. The latter information  confirms  the qualitative relation between SF and exchange cycles involving molecules in different spatial regions of the cluster.   
Clearly,   the local superfluid response of this cluster  is never confined to the surface. It  is important to note that the discussed variations of $\rho_S(r)$ occur against  a total density profile which is essentially unaffected by temperature, for this liquidlike cluster (as shown in Fig. \ref{fig:18}). 

\begin{figure}
\centerline{\includegraphics[scale=0.37,angle=-90]{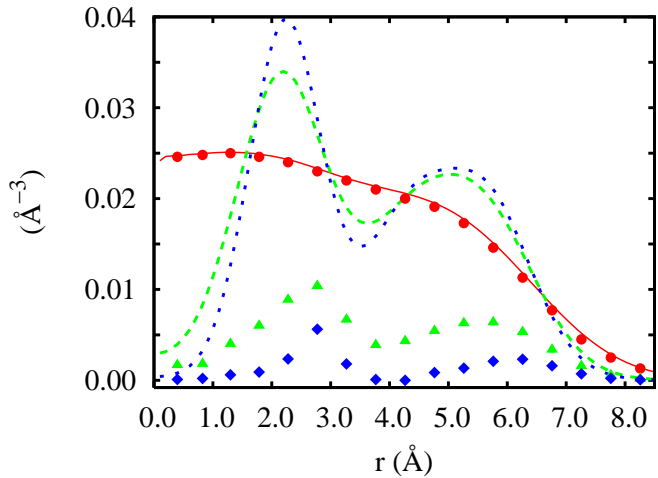}}
\caption{(color online) Profiles of total  and superfluid density at $T$ = 0.0625 K (solid line and circles), $T$ = 0.25 K (dashed line and triangles) and $T$ =0.5 K (dotted line and diamonds) for the cluster  (\ph2)$_{26}$.}
\label{fig:26}
\end{figure}

A behavior analogous to that illustrated above for the (\ph2)$_{18}$ cluster is displayed essentially by all clusters with fewer than 22 molecules, which are liquidlike in the temperature range considered here. As discussed in Refs. \cite{noi06,noi07a}, clusters with a number of molecules between 22 and 30 feature a different physical behavior, displaying (to different degrees) coexistence of liquid- and solidlike properties at $T\le$ 1 K. We discuss here the specific case (\ph2)$_{26}$.

%NOTA: cambiata leggermente la discussione della figura
Figure \ref{fig:26}  shows the same profiles as Fig. \ref{fig:18} but  for a cluster of $N$=26 molecules. This cluster displays  insulating properties at $T$=0.5 K; its superfluid fraction is barely over  10\%, rising to approximately 32\% at $T$ = 0.25 K and reaching 100\% at $T$ = 0.0625 K. Correspondingly, and quite unlike the $N$=18 case, the structure of the cluster undergoes a significant change, as shown by the evolution of the radial density profile. In particular, the peak corresponding to the first shell (at r $\simeq$ 2 \AA) becomes depressed  as $T$ decreases; hardly any remnant of the shell structure is left at $T$ = 0.0625 K, as a liquidlike structure emerges, i.e.,  the cluster undergoes quantum melting \cite{noi06, noi07a}. Indeed, it is at low $T$ (in contrast with what suggested in Ref. \cite{buffoni}), when this cluster quantum melts, that its basic physics becomes qualitatively reminiscent of that of a helium cluster.

\begin{figure}[h]
\centerline{\includegraphics[scale=0.37,angle=-90]{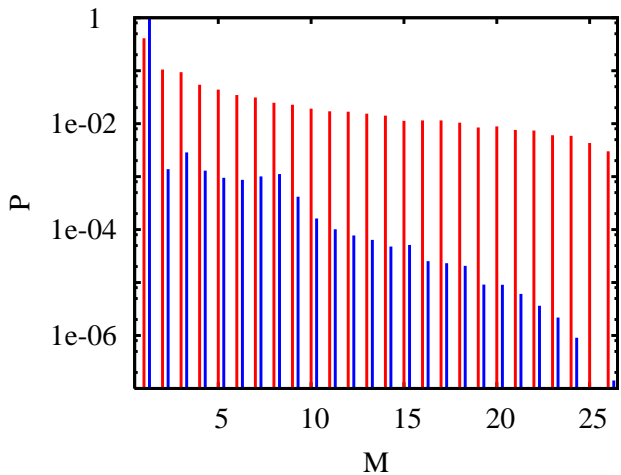}}
\caption{(color online) Frequency of occurrence of permutation cycles involving $M$ molecules in a cluster of 26 \ph2 molecules at $T$=0.5 K (lower values) and $T$=0.0625 K (higher values). Long permutation cycles set in at low $T$.}
\label{fig:prm}
\end{figure}

It is interesting to examine the corresponding evolution of the local superfluid density, also shown in Fig. \ref{fig:26}.
At the highest temperature shown, $\rho_S(r)$  is small at all distances from the center of mass, and negligible  between the two shells (diamonds); here too it has to be stressed  how, though the superfluid response is fractionally greater in the immediate surface (i.e., $r > 7$ \AA), it is nonetheless not confined thereto. Moreover, as the temperature is lowered and the cluster turns liquidlike and superfluid,  the relative increase of $\rho_S(r)$  is far more significant in its inner part. Especially telling is the disappearance of the dip in the region between the two shells (at $r\sim 4$ \AA), suggesting the occurrence of permutations involving particles in both the inner and outer shell. 
This is consistent with the observed temperature dependence of the frequency of occurrence of permutation cycles involving different numbers of molecules, shown in Fig. \ref{fig:prm}.

We discussed in previous work \cite{noi06, noi07a} the non-monotonic behavior of the total superfluid fraction of clusters of size $N \geq 22$; we further illustrate this phenomenon here, by focusing on a very narrow range of cluster sizes, within which the value of the superfluid fraction changes dramatically on adding just one molecule at a time.
The results for the local superfluid, and total density definitively demonstrate that such behavior reflects structural changes throughout the {\it whole} cluster, not just its surface as contended in Ref. \cite{buffoni}. 
\begin{figure}
\centerline{\includegraphics[scale=0.37,angle=-90]{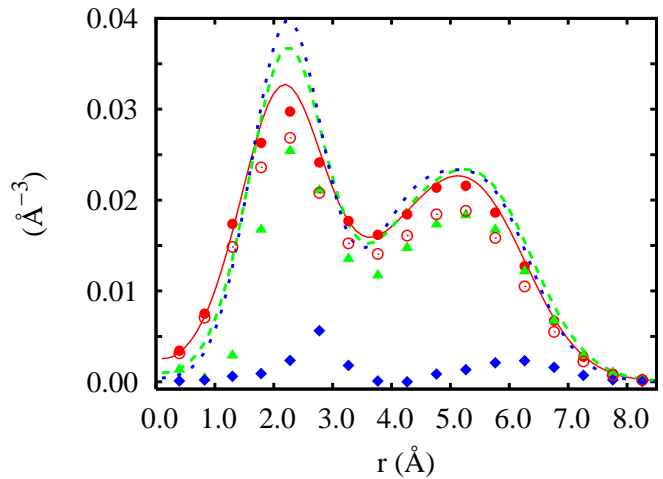}}
\caption{(color online) Profiles of total  and superfluid density for the cluster (\ph2)$_{25}$ (solid line and circles), (\ph2)$_{26}$ (dotted line and diamonds),(\ph2)$_{27}$ (dashed line and triangles) at $T$ = 0.5 K. Open symbols show estimates of the local superfluid density obtained using the same measure utilized in Ref. \onlinecite{buffoni}.}
\label{fig:size}
\end{figure}

Total and superfluid densities for clusters of 25, 26 and 27 molecules at a temperature of 0.5 K are shown in Fig. \ref{fig:size}.  The superfluid fraction is close to 1 for $N$ = 25, dropping to roughly 0.12 for $N$=26 and rising again up to approximately 0.85 for $N$=27. For the $N$=25 cluster, the superfluid response is fractionally close to 100\% throughout the entire cluster. While obviously fewer particles reside on average in the core than on the surface region of such a small cluster,  the results of Fig. \ref{fig:size}  again clearly show how SF is  not confined to the surface, and the core of the cluster is not rigid but superfluid. The addition of a single molecule causes the entire cluster to turn abruptly insulating and solidlike, as shown by the increase in the height of the total density first shell peak.
The local superfluid response is concurrently depressed throughout the whole system, especially in the region between the two shells, to indicate the paucity of exchanges taking place between molecules in different shells. On adding yet another molecule, i.e., on increasing the cluster size to $N$=27, the cluster returns to a liquidlike structure, and consistently the superfluid response is enhanced, at the surface, but also in  the inner part of the system. 
These findings lend substantial theoretical support to the notion that the liquidlike, superfluid phase of these small clusters occurs as a result of the onset of long exchanges involving all molecules in the system. This is in fundamental disaccord with the picture put forth in Ref. \cite{buffoni}, according to which SF would correlate to long exchanges involving loosely bound surface molecules. 

%It seems reasonable to attribute the discrepancy of our results  with those of Ref. \cite{buffoni} mainly  to the estimator for the local superfluid density. The estimator adopted here, rigorously derived in Ref. \cite{whaley}, differs substantially from the one employed in Ref. \cite{buffoni}, which, to our knowledge, has no real theoretical basis. The difference between the results furnished by the various measures of the local superfluid density adopted in previous works has been extensively  discussed in Ref. \cite{whaley}.
One might be tempted to attribute the discrepancy of our results with those of Ref. \cite{buffoni} to the   difference between the estimator of the local superfluid density utilized here, and the qualitative measure  employed in Ref. \cite{buffoni}, which lacks any microscopic justification \cite{whaley,draeger}.  For comparison, we show in Fig. \ref{fig:size} a superfluid density profile obtained using the same measure of Ref. \cite{buffoni} (open circles); no substantial difference is seen with respect to our result (full circles). Furthermore, our physical conclusions can also be inferred from the evolution of the density profiles as the temperature is lowered (Fig. \ref{fig:26}), clearly showing the disappearance of the shell structure  and the concurrent approach to unity of the {\it total} superfluid fraction (for this quantity the same estimator is used in our work and Ref. 4), as well as by the temperature dependence of the frequency of occurrence of long permutation cycles (Fig. \ref{fig:prm}). All of this unambiguously shows  that superfluidity correlates 
with long exchanges, comprising molecules both in the inner and outer shells. 

More generally, it should be pointed out that the continuous-space Worm Algorithm, utilized in this work enjoys \cite{MBworm,worm2} a much greater efficiency in sampling permutations than the method used in Ref. \onlinecite{buffoni}. It is quite conceivable that
the physical behavior observed in this work, underlain by long permutation cycles,  may have remained hidden to the authors of Ref. \onlinecite{buffoni}, due to the inefficiency of their method in sampling such cycles, especially at the lowest temperatures.

In order to explore a possible dependence of the physical results on the potential utilized,  we have also  performed in this work calculations with the Buck potential \cite{buck} (another popular choice), as well as with the Lennard-Jones potential utilized in Ref. \cite{buffoni}. Our main  physical conclusions are unaffected by the intermolecular potential utilized, which essentially only influences the energetics and the temperature scale.

Summarizing we have studied the size and temperature  dependence of the  local superfluid density of \ph2 clusters. We found no evidence of superfluid response confined at the surface,  regardless of  temperature and system size explored. SF concomitantly grows as $T$ decreases in the inner and outer part of the system in liquidlike clusters [i.e., (\ph2)$_{18}$] as well as in those which feature quantum melting at low $T$. In the latter case,  structural variations  such as the progressive collapse of the first peak of the radial density profile are due to long (i.e., inter-shell) exchanges, involving all molecules in the cluster.  Long exchanges cause the system to turn liquidlike at sufficiently low $T$, with the ensuing onset of superfluidity. Significant superfluid response is only seen when the the core and the surface are both superfluid and liquidlike, i.e., it is not  the effect  of quantum exchanges  involving exclusively, or mostly, surface molecules. 
%A more general conclusion is that in the low temperature limit, as they become increasingly liquidlike, %the physics of these small parahydrogen clusters does not seem qualitatively different than that of %helium clusters.

This work was supported  by the Natural Science and Engineering Research Council of Canada under research grant 121210893, and by the Informatics Circle of Research Excellence (iCORE).
Simulations were performed on the Mammouth cluster at University of Sherbrooke (Qu\'ebec, Canada).

\end{document}